\documentclass[aps,prd,twocolumn,superscriptaddress,showpacs,nofootinbib]{revtex4-1}

\usepackage{epsfig}
\usepackage{graphicx}% Include figure files
\usepackage{dcolumn}% Align table columns on decimal point
\usepackage{bm}% bold math
\usepackage{multirow}
\usepackage{array}
\usepackage{slashed}
\usepackage{amsmath}
\usepackage[titletoc]{appendix}
\usepackage{color}
\usepackage{hyperref}
\usepackage{soul}

\begin{document}

\title{Effects of Annihilation with Low-Energy Neutrinos on High-Energy Neutrinos from Binary Neutron Star Mergers and Rare Core-Collapse Supernovae} 

\author{Gang Guo}
%\email{guogang@cug.edu.cn} 
\affiliation{School of Mathematics and Physics, China University of Geosciences, Wuhan 430074, China}
\author{Yong-Zhong Qian}
%\email{qian@physics.umn.edu} 
\affiliation{School of Physics and Astronomy, University of Minnesota, Minneapolis, MN 55455, USA}
\author{Meng-Ru Wu}
%\email{mwu@gate.sinica.edu.tw}
\affiliation{Institute of Physics, Academia Sinica, Taipei, 11529, Taiwan}
\affiliation{Institute of Astronomy and Astrophysics, Academia Sinica, Taipei, 10617, Taiwan}
\affiliation{Physics Division, National Center for Theoretical Sciences, Taipei 10617, Taiwan}

\date{\today}

\begin{abstract}
We explore the possibility that high-energy (HE) neutrinos produced from choked jets
can be annihilated with low-energy (LE) neutrinos emitted from the accretion 
disk around a black hole in binary neutron star mergers and rare core-collapse supernovae. For HE neutrinos produced close
to the stellar center ($\lesssim 10^{9}$--$10^{12}$ cm), we find that the emerging all-flavor spectrum for neutrinos of $E\gtrsim 0.1$--1 PeV could be modified by a factor $E^{-n}$ with $n\gtrsim 0.4$--0.5 under realistic conditions. Flavor evolution of LE neutrinos does not affect this result but can change the emerging flavor composition of HE neutrinos. As a consequence, the above annihilation effect may need to be considered for HE neutrinos produced
from choked jets at small radii. We briefly discuss the annihilation effects for different HE neutrino production models and point out that such effects could be tested through precise measurements of the diffuse neutrino spectrum and flavor composition.
\end{abstract}

%\pacs{95.85.Ry, 97.60.Bw, 98.70.Rz, 14.60.Pq} 

\maketitle

\section{introduction}
The detection of cosmic high-energy (HE) neutrinos by IceCube \cite{Aartsen2013a,Aartsen2013b,Aartsen2014,Aartsen2015} 
ushered in a new era of neutrino astronomy. While some IceCube events are correlated with the blazar TXS 0506+056 
\cite{Aartsen2018,Ackermann2018}, three tidal disrupted events~\cite{Stein:2020xhk,Reusch:2021ztx,vanVelzen:2021zsm}, 
and an active galaxy~\cite{IceCube:2022der}, the sources of TeV--PeV neutrinos remain mostly unidentified. 
Nevertheless, there are many constraints on contributions from various sources and the associated models. 
For instance, lack of direction and time correlation with gamma-ray bursts (GRBs) \cite{Abbasi2012_grb,Aartsen2015_grb,Aartsen2016_grb,Aartsen2017_grb,IceCube:2022rlk} 
limited their contribution to be less than 1\%\footnote{The nondetection of HE neutrinos from the brightest GRB, GRB 221009A, 
sets a limit comparable to that from the stacked searches \cite{Murase:2022vqf,Ai:2022kvd,Liu:2022mqe,Rudolph:2022dky,IceCube:2023rhf}.} 
and hence led to reconsideration of HE neutrino production in GRBs \cite{Hummer:2011ms,Li_2012,Zhang:2012qy}.
Because both HE neutrinos and $\gamma$-rays are produced from meson decay, sources are likely opaque to $\gamma$-rays 
in order to avoid overproducing the observed diffuse $\gamma$-ray background \cite{Murase2013_pp,Murase2016,Murase:2019vdl,Capanema:2020rjj,Fang:2022trf}.
While current data allow a wide range of flavor composition, sources with the standard pion or neutron decay scenarios are constrained
\cite{Mena2014,Aartsen2015_sp,Aartsen2015_fl,Palladino2019,Bustamante2019}. 

Many other issues regarding HE neutrino production and propagation are worth exploring.
For example, in specific astrophysical environments, microscopic processes beyond those included in the current models may be important.
New physics beyond the standard model such as non-standard interaction of neutrinos can alter the spectrum and flavor content of 
the HE neutrinos reaching the Earth \cite{Beacom2002,Pakvasa2012,Shoemaker2015,Moss2017,Denton2018,Moulai:2019gpi,Abdullahi:2020rge}. 
By accumulating statistics on the diffuse flux, IceCube, and especially IceCube-Gen2, can provide even better probes of the sources and 
related neutrino physics \cite{Ackermann:2017pja,IceCube-Gen2:2020qha,Song:2020nfh,Kochocki:2023lhh}. Detection of many events from 
a single nearby source would also give powerful constraints. With the above considerations in mind, here we discuss low-energy 
(LE) neutrino emission associated with the central engines of GRBs and rare core-collapse supernovae (CCSNe) that produce HE neutrinos, and
explore how annihilation with these LE neutrinos may affect the spectrum and flavor composition of the HE neutrinos emerging from these sources.

Short GRBs lasting $\sim 0.1$--1 s are associated with binary neutron star mergers (BNSMs) while long GRBs lasting a few seconds are mostly associated with
rare CCSNe, the so-called collapsars or hypernovae. In both cases, an accreting black hole is widely considered as one of the primary candidates 
for the central engine of GRBs \cite{Woosley1993}. The accretion disk associated with the black hole can emit profuse fluxes of nearly thermal LE
neutrinos of ${\cal O}(10)$ MeV, mainly $\nu_e$ and $\bar\nu_e$. 
Relativistic jets may be powered either by annihilation of these LE $\nu_e$ and $\bar\nu_e$ or by extracting the rotational energy of 
the black hole through the Blandford-Znajek mechanism \cite{Popham1999}. Shocks can occur at different stages of jet propagation, 
leading to different scenarios for HE neutrino production\footnote{The internal and external shocks, which are responsible for 
the prompt radiation and afterglow of GRBs, respectively, can generate prompt neutrinos \cite{Waxman1997,Guetta:2003wi,Murase:2006mm,Hummer:2011ms,Li_2012,Zhang:2012qy,Liu2013,Bustamante2015,Biehl:2017zlw} 
and neutrino afterglow \cite{Waxman:1999ai,Dermer2000,Dai2001,Waxman2002,Li2002,Murase:2007yt,Razzaque2013,Razzaque2014,Nir:2015teo,Thomas2017}. 
Furthermore, HE neutrinos associated with X-ray flares and late rebrightening, which cannot be well explained by the standard afterglow theory, 
have also been investigated \cite{Murase:2006dr,Guo:2019ljp}.} \cite{Kimura:2022zyg}. As the jet propagates through the ejecta from a BNSM or 
the envelope of a CCSN, HE neutrinos can be produced at internal shocks before jet collimation, at collimation shocks, and 
at forward and reverse shocks driven by the mildly-relativistic or non-relativistic jet head 
\cite{Meszaros2001,Razzaque:2003uv,Razzaque2004,Ando:2005xi,Razzaque2005,Horiuchi:2007xi,Enberg:2008jm,Bartos:2012sg,Murase2013,Fraija2013,Xiao2014,Bhattacharya2014,Varela2015,Tamborra2015,Senno2015,Fraija2015,Senno:2017vtd,Denton:2017jwk,Denton:2018tdj,He:2018lwb,Gottlieb:2021pzr,Chang:2022hqj,Guarini:2022hry,Bhattacharya:2022btx,Reynoso:2023qoh}. 
In the case of rare CCSNe, it may be more common that the jet is not energetic enough to penetrate the whole stellar envelope, and
the resulting choked jet is dark in electromagnetic radiation but bright in HE neutrinos. In addition, both successful and failed jets 
may transfer energy to stellar matter, driving a mildly-relativistic cocoon that produces low-luminosity GRBs \cite{Murase2013,Gottlieb2018} 
accompanied by HE neutrinos during shock breakout \cite{Waxman2001,Katz2011,Kashiyama2013}. If the jets are magnetically dominated, magnetic reconnection can act as another mechanism for particle acceleration \cite{McKinney_2011,Zhang:2011,Guo:2016}. Because we are interested in HE neutrinos and
their annihilation with LE neutrinos, we focus on those BNSMs and rare CCSNe that can produce both types of neutrinos regardless of any associated GRBs.

The impact of annihilation with LE neutrinos on HE neutrinos received little attention in previous studies. 
The only exception is our recent study \cite{Guo:2022zyl}, where we considered collapsars as sources for both HE neutrinos and $r$-process nuclei.
The $r$-process nuclei are synthesized in the nonrelativistic winds from the accretion disk \cite{Wei:2019hpd,Chen:2023mn,An:2023edd}.
The $\beta$-decay of these radioactive nuclei produces LE $\bar\nu_e$, which first oscillate into $\bar\nu_\mu$ and then can annihilate HE $\nu_\mu$
produced by shocks associated with jet propagation. We demonstrated that such annihilation could leave imprints on the spectrum and flavor composition 
of the emerging HE neutrino flux \cite{Guo:2022zyl}. In this work, we conduct a similar study, but as stated above,
our focus is on the LE neutrinos emitted by the accretion disk.
Given the complexity of HE neutrino production in different sites and the extensive studies already available in the literature, 
we choose to present a largely parametric study on how annihilation with LE neutrinos affects the emerging HE neutrino flux. 
Our results can be used to assess the significance of such annihilation for specific scenarios of HE neutrino production.

This paper is organized as follows. In Sec. \ref{sec:adnu} we discuss emission of LE neutrinos by 
an accretion disk in BNSMs or rare CCSNe and the flavor evolution of both LE and HE neutrinos.
Without addressing the detailed mechanism of HE neutrino production, we present in Sec. \ref{sec:anni} 
a parametric study on how annihilation with LE neutrinos may affect HE neutrinos.
We then  briefly discuss such effects on HE neutrinos produced by choked jets from BNSMs and CCSNe within various models, and conclude in Sec. \ref{sec:summary}.

\section{emission of LE neutrinos and flavor evolution of LE and HE neutrinos} 
\label{sec:adnu}  

\subsection{Luminosity and duration}      

Abundant fluxes of MeV-scale 
$\nu_e$ and $\bar\nu_e$ can be emitted from accretion disks mainly via $e^\pm$ captures on nucleons. For hyperaccreting rate of ${\dot M} \sim$ 0.001--10 $M_\odot/s$, the fraction of energy converted to neutrinos, $\epsilon_\nu\equiv L_\nu/\dot Mc^2$, can vary widely from 
$\sim 0.01$--$0.3$, depending on the initial mass and spin of the black hole, 
the accretion rates, and the viscosity of accretion disk \cite{Popham1999,Di_Matteo2002,Chen2007,Liu2007,Shapiro:2017cny}. Generally, lower viscosity and higher black hole spin give rise to higher values of $\epsilon_\nu$. For $\dot M\gtrsim 0.1 M_\odot$/s, $\epsilon_\nu$ can reach 0.05 for both a non-rotating (Schwarzschild) and a fast rotating (Kerr) black hole \cite{Chen2007,Just2015}. The corresponding neutrino luminosity can be as high as $10^{53}$ erg/s for $\dot M = 1 M_\odot$/s, and even higher for accretion rates of 10 $M_\odot$/s, which could occur in the case of compact object mergers \cite{Chen2007,Schilbach:2018bsg}. The jet luminosity is related to $L_\nu$ if the jet is powered by pair annihilation of accretion disk neutrinos. Previous studies indicated that the pair annihilation luminosity $L_{\bar\nu\nu} \sim (10^{-3}$--$10^{-1}) L_\nu$ \cite{Popham1999,Di_Matteo2002,Liu2007}, which is consistent with the observed luminosity for classical GRBs with $L_{\rm ob} \sim 10^{51}$~erg/s. The mean energies of accretion disk neutrinos are typically $\sim 10$--20 MeV \cite{Just2015}. The duration of the accretion process and neutrino emission could be similar to that of the associated GRBs, which are $\sim 0.1$--1 s and $\sim$ a few seconds for short \cite{Just2015} and long GRBs \cite{Wei:2019hpd}, respectively.

Instead of using any specific numerical models, we simply assume that the neutrino luminosities from
the accretion disk
are constant over time, and choose a standard Fermi-Dirac distribution
with the same effective $T_\nu$ and zero chemical potential for $\nu_e$ and $\bar\nu_e$ to describe the spectra. For thermal neutrinos emitted from a surface area of $S \sim 2\pi R_\nu^2$ with an effective emission radius $R_\nu = 10^7$ cm, we have $L_\nu \sim \sigma T_\nu^4  S$, with $\sigma$ being the Stefan-Boltzmann constant. For $T_\nu = 5$ MeV, the corresponding luminosity is $L_\nu \sim 10^{53}$ erg/s.
Without referring to 
specific models, we vary $T_\nu$ in the range  
from 3 to 8 MeV for the parametric study. Correspondingly, the total neutrino luminosity varies from $10^{52}$ to $10^{54}$~erg/s, which is broadly consistent with numerical simulations~\cite{Cusinato:2021zin,Fernandez:2023vxj,Just:2022fbf,Fujibayashi:2022xsm}.

It is crucial to consider the contemporaneity of LE neutrino emission by the accretion disk and HE neutrino production by the jets. Shortly after the onset of thermal neutrino emission, jets are launched immediately from
the accretion disk.
With the jet velocity $\beta_j c$ and Lorentz factor $\Gamma_j=(1-\beta_j^2)^{-1/2}$, the time lag $\Delta t$ for the jets to reach
the shock formation site at radius $R_\mathrm{sh}$ relative to the thermal neutrinos is $\sim R_{\rm sh} /(2\Gamma_j^2c)$ for mildly relativistic or relativistic jets, and $\sim R_{\rm sh}/(\beta_j c)$ for non-relativistic jets. To ensure that the HE neutrinos meet the LE neutrinos,
the time lag needs to be smaller than 
the duration of thermal neutrino emission $\Delta T$. In the case of mildly relativistic or relativistic jets, this indicates that $R_{\rm sh} \lesssim 6 \times 10^{11}(\Gamma_j/3)^2 (\Delta T/{\rm s})$~cm. For nonrelativistic jets, $R_{\rm sh}$ should be smaller than $\sim 3 \times 10^{10}\beta_j (\Delta T/{\rm s})$~cm.
Our study focuses on cases where HE neutrinos are produced close to the center at $R_{\rm sh} \sim 10^{9}$--$10^{12}$ cm so that a significant impact of the LE neutrinos could be expected [see, e.g., Eq. (\ref{eq:tau2}) below]. Given the above considerations, LE and HE neutrinos can meet for annihilation except for cases with $\beta_j \ll 1$ and $\Delta T \ll 1$ s.

\subsection{Flavor evolution of LE and HE neutrinos}  

For studying their effects on jet-produced   
HE neutrinos, the flavor evolution of the thermal neutrinos due to different mechanisms, 
including the collective oscillations~\cite{Malkus2012,Duan2010,Wu:2017qpc,Fischer:2023ebq} and the Mikheyev-Smirnov-Wolfenstein (MSW) effect 
\cite{Wolfenstein1978,MikheyevS.P.andSmirnov1985}, is a crucial input.
For simplicity, we assume that only $\nu_e$ and $\bar\nu_e$ \cite{Popham1999,Malkus2012} are emitted from the accretion disk, and the probability for an initial $\nu_e$ ($\bar\nu_e$) to become a $\nu_\beta$ ($\bar\nu_\beta$; $\beta=e$, $\mu$, $\tau$) at radius $r\gg R_\nu$ outside the accretion disk is parametrized as $f_\beta(r)$ [$\bar f_\beta(r)$]. 
Instead of solving the detailed flavor evolution of thermal neutrinos, we consider the following five different flavor evolution scenarios for accretion disk neutrinos (see Tab.~\ref{tab:flavor_evol}):
(1) no evolution (NE), for which $f_\beta(r)=\bar f_\beta(r)=\delta_{\beta e}$ 
in the Kronecker $\delta$ notation, 
(2) adiabatic evolution with normal mass ordering (NO), for which 
$f_\beta(r)=|U_{\beta3}|^2$ and $\bar f_\beta(r)=|\bar U_{\beta1}|^2$
in terms of vacuum mixing matrix elements $U_{\beta i}$ and 
$\bar U_{\beta i}$ ($i=1$, 2, 3),
(3) adiabatic evolution with inverted mass ordering (IO), for which 
$f_\beta(r)=|U_{\beta2}|^2$ and $\bar f_\beta(r)=|\bar U_{\beta3}|^2$,
(4) exotic evolution (EE), for which $f_\beta(r)=\bar f_\beta(r)=\delta_{\beta\mu}$, and (5) flavor equipartition (FE) with $f_\beta=f_{\bar\beta}=1/3$.
We use appropriate best-fit values of mixing parameters from \cite{Gonzalez-Garcia2014} to evaluate $U_{\beta i}$ and $\bar U_{\beta i}$.

\begin{table*}[htbp] 
\centering 
\caption{Flavor evolution scenarios for LE neutrinos.}
\renewcommand{\arraystretch}{1.5}
\begin{tabular}{p{2.8cm}<{\centering}p{3.7cm}<{\centering}p{3.7cm}<{\centering}p{3.2cm}<{\centering}p{3.5cm}<{\centering}}    
\hline
\hline
no evolution & adiabatic evolution assuming NO & adiabatic evolution assuming IO & exotic evolution & flavor equipartition \\
 (NE) & (NO) & (IO) & (EE) & (FE)  \\     
\hline
  $f_\beta(r)=\bar f_\beta(r)=\delta_{\beta e}$ & $f_\beta(r)=|U_{\beta2}|^2$, $\bar f_\beta(r)=|\bar U_{\beta3}|^2$ &
$f_\beta(r)=|U_{\beta2}|^2$, $\bar f_\beta(r)=|\bar U_{\beta3}|^2$ & $f_\beta(r)=\bar f_\beta(r)=\delta_{\beta\mu}$ & $f_\beta=f_{\bar\beta}=1/3$ \\
\hline
\end{tabular}
\label{tab:flavor_evol} 
\end{table*}

The above five scenarios may be realized in CCSNe and BNSMs under different physical conditions. 
For a neutrino of energy $E_\nu$, the MSW effect
takes place at 
a resonance density $\rho_{{\rm res},7}\approx 1.3E_{\nu,{\rm MeV}}^{-1}
(\delta m^2/{\rm eV}^2)\cos2\theta_v$, where $\delta m^2$ is the 
vacuum-mass-squared difference and $\theta_v$ is the vacuum mixing angle.
Here and below, we use subscripts to indicate eV-based units and powers of 10 in cgs units, i.e., $a_x \equiv a/10^x$.
For accretion disk neutrinos of $E_\nu\sim 10$~MeV, two resonances occur at high and
low densities of $\rho_{H,3}\sim 3$ and $\rho_{L,1}\sim 4$ for 
$(\delta m^2/{\rm eV}^2,\theta_v)=(2.4\times10^{-3},8.5^\circ)$ and
$(7.5\times 10^{-5},33.5^\circ)$, respectively \cite{Gonzalez-Garcia2014}.
As $\rho\gg\rho_H$ at the accretion disk, both resonances are relevant 
if $\rho<\rho_{\rm L}$ at $r\sim R_{\rm sh}$.
Such a condition can be fulfilled for CCSNe with $R_{{\rm sh},10}\gtrsim 3$ 
\cite{Varela2015} and for BNSMs \cite{Aloy2005}.
Assuming that the collective oscillations are neglected and the flavor evolution through both resonances are adiabatic, this 
corresponds to the scenario NO or IO, depending on the yet-unknown
neutrino mass ordering.
For the scenario NE, it may occur in CCSNe with $R_{{\rm sh},9}\sim 3$, where the accretion disk neutrinos do not go through MSW 
resonances before reaching $r\sim R_{\rm sh}$ with $\rho>\rho_H$ \cite{Varela2015} when collective oscillations are ignored.
For the collective oscillations expected to occur near the accretion disk where the neutrino densities are high~\cite{Malkus2012,Duan2010,Wu:2017qpc,Fischer:2023ebq},
the effect can be complicated and is under intense investigation. 
If it happens, it will lead to different flavor evolution history from the above three scenarios, and we use the scenarios EE and FE to represent the range of possible outcomes.
In short, the exact flavor evolution of accretion disk neutrinos for specific models requires more detailed treatment but
we expect 
the outcome to be within the range covered by the above five representative scenarios.

For the HE neutrinos, the adiabaticity is typically broken. As a consequence, HE neutrinos with $E_\nu \sim 1$--30 TeV propagating inside the stars or the ejecta from the BNSMs can undergo substantial nonadiabatic flavor conversion \cite{Mena:2006eq,Razzaque:2009kq,OsorioOliveros:2013azu,Xiao:2015gea,Carpio:2020app}.
The resulting flavor composition of HE neutrinos at sites relevant for neutrino pair annihilation depends on both the initial flavor composition produced and the density profile that HE neutrinos traverse.
For even higher energies, the flavor evolution of HE neutrinos could be neglected (see Ref.~\cite{Abbar:2022hgh} for discussion on how LE neutrinos might induce flavor evolution of HE neutrinos). Given our focus on the effects of neutrino pair annihilation in this study, we choose not to take into account the flavor evolution of HE neutrinos in discussing their annihilation with LE accretion disk neutrinos. It is also worth noting that, under realistic conditions, only neutrinos with $E_\nu \gtrsim 10$ TeV can be significantly affected by pair annihilation, as demonstrated later.

\section{$\bar\nu\nu$ annihilation of LE and HE neutrinos}   
\label{sec:anni}

\begin{figure}[htbp]
\includegraphics[width=\columnwidth]{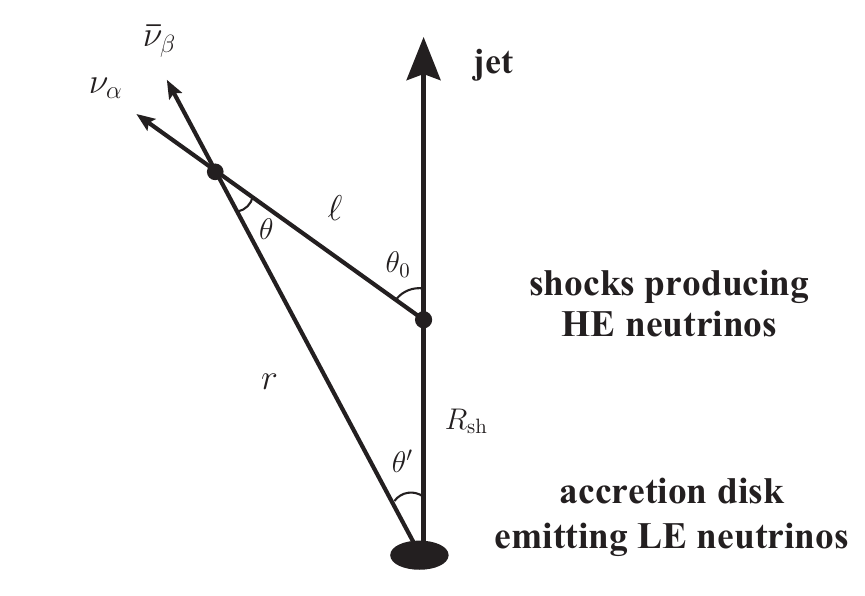}
\caption{Sketch of interaction between HE $\nu_\alpha$ and LE $\bar\nu_\beta$.}
\label{fig:Sketch}
\end{figure}

In this section we study how the flux and flavor composition of HE neutrinos are modified by annihilation with the LE neutrinos.
Without referring to specific models, we simply assume that the HE neutrinos are produced at a radius $R_{\rm sh}$ representative of shocks accelerating protons\footnote{In addition to shock acceleration, protons can also be accelerated through the neutron-proton-converter mechanism \cite{Murase:2013hh,Kashiyama:2013ata} or magnetic reconnections. Note that we focus on the shock case but our parametric study applies equally to the other cases.}. Consider a HE  
$\nu_\alpha$ of energy $E$ emitted with an  
angle $\theta_0$ relative to the jet propagation direction
(Fig.~\ref{fig:Sketch}) and assume that the HE neutrino can always meet the thermal neutrinos along its trajectory. 
When it interacts with an LE $\bar\nu_\beta$ of energy $E'$ 
at radius $r$, the main processes are
\begin{equation}
\nu_\alpha\bar\nu_\beta\to\left\{\begin{array}{ll}
f\bar f,&\alpha=\beta,\\
l_\alpha\bar l_\beta,&\alpha\neq\beta,
\end{array}\right.
\end{equation}
where $f$ stands for the relevant quarks and leptons and $l$ for the 
charged leptons. The corresponding cross sections 
$\sigma_{\nu_\alpha\bar\nu_\beta}(s)$ \cite{Roulet1993} are functions of 
$s=2EE'(1-\cos\theta)$, where 
$\theta$ is the intersection angle between $\nu_\alpha$ and $\bar\nu_\beta$.
The probability for the HE $\nu_\alpha$ to survive annihilation,
$P_{\nu_\alpha}(E,\theta_0)=\exp[-\tau_{\nu_\alpha}(E,\theta_0)]$, is
determined by the ``optical'' depth
\begin{equation}
\tau_{\nu_\alpha}(E,\theta_0)=\sum_\beta\int(1-\cos\theta)
\sigma_{\nu_\alpha\bar\nu_\beta}(s)dn_{\bar\nu_\beta}(E', r)d\ell, 
\label{eq:tau}
\end{equation}
where $\ell$ is the path length of $\nu_\alpha$, 
\begin{equation}
dn_{\bar\nu_\beta}(E', r)=\frac{E'^2dE'}{\exp(E'/T_\nu) + 1}
\frac{R_\nu^2\cos\theta'}{8\pi^2r^2}\bar f_\beta(r),
\label{eq:nnu}
\end{equation}
is the energy-differential number density of $\bar\nu_\beta$ 
at radius $r$, and $\theta'=\theta_0-\theta$. 
Note that $\theta$ and $r$ can be solved from $R_{\rm sh}$, $\theta_0$, 
and $\ell$ (Fig.~\ref{fig:Sketch}).

Taking $\theta\sim\theta_0\ll 1$, $\ell\sim r\sim R_{\rm sh}$, and
$\sigma_{\nu_\alpha\bar\nu_\beta}\sim G_F^2s$, where $G_F$ is the Fermi coupling constant, we can estimate
\begin{eqnarray}
\tau_{\nu_\alpha}(E,\theta_0)&\sim&\frac{7\pi^2}{1920}G_F^2E
\frac{R_\nu^2T_\nu^4\theta_0^4}{R_{\rm sh}}\nonumber\\
&\sim&25E_{\rm PeV}
R_{\nu,7}^2T_{\nu,{\rm MeV}}^4\theta_0^4R_{{\rm sh},9}^{-1}.
\label{eq:tau2}
\end{eqnarray}

The above estimate shows that 
$P_{\nu_\alpha}(E,\theta_0)$ is sensitive to the emission angle
$\theta_0$ of $\nu_\alpha$ relative to the jet propagation direction. In general, the typical value of $\theta_0$ depends on the jet geometry during propagation. We first consider a simple scenario where jets with a finite opening angle move along the radial direction (with the black hole sitting at the center).
This scenario includes conic jets initially launched from the center before being collimated, or collimated jets influenced by stellar matter. Assuming that HE neutrinos are emitted isotropically in the rest frame of the shocked fluid at radius $R_{\rm sh}$, the corresponding normalized distribution of $\theta_0$ [$\int_{-1}^1 g_\Gamma(\theta_0)d\cos\theta_0=1$] can be obtained after a Lorentz boost to the stellar rest frame as:
\begin{align}
g_\Gamma(\theta_0) = {1-v^2 \over 2(1-v\cos\theta_0)^2},
\label{eq:gem}
\end{align}
where $v$ is the bulk velocity of the shocked fluid with $\Gamma_\mathrm{sh}\equiv(1-v^2)^{-1/2}$ and $\Gamma_\mathrm{sh}$ is the Lorentz factor as observed in the stellar rest frame. Since the typical value of $\theta_0$ is of order $\Gamma_{\rm sh}^{-1}$ for this scenario, we expect that only HE neutrinos emitted from mildly relativistic jets (with $\Gamma_{\rm sh} \lesssim 10$; see below) can undergo a significant annihilation effect.

The interaction with stellar matter or ejecta from BNSMs could impact the jet structure and dynamics, causing them to deviate from a strictly radial trajectory. Depending on where HE neutrinos are produced and emitted, the values of $\theta_0$ could avoid the constraint set by the $\Gamma_{\rm sh}^{-1}$, allowing HE neutrinos emitted from even ultra-relativistic shocked fluid to experience substantial annihilation with LE neutrinos. A notable example occurs when jets, initially collimated, would undergo a rapid lateral expansion upon breaking out of the stellar matter or the ejecta from a BNSM, reaching an opening angle ranging from a few to tens of degrees \cite{Lazzati:2005xv,Mizuta:2013yma}. If HE neutrinos are generated by jets after breakout, the corresponding values of $\theta_0$ would be of the order of the jet opening angle, regardless of $\Gamma_{\rm sh}$. A similar situation arises when the shock driven by the jet and cocoon breaks out, transforming into a collisionless shock capable of accelerating protons and producing HE neutrinos \cite{Senno2015,Gottlieb2018,Gottlieb:2021pzr}. Typically, this shock can span an opening angle considerably wider than that of the jets. For the scenario discussed above, we simply consider a normalized distribution: 
\begin{align}
g_\theta(\theta_0) = {1 \over 1-\cos\theta_{\rm sh}},\ \theta_0\leq\theta_{\rm sh},\label{eq:gem_theta}
\end{align}
which corresponds to a uniform distribution of $\cos\theta_0$ with $\theta_0\leq\theta_{\rm sh}$.

\begin{figure}[htbp]
\includegraphics[width=\columnwidth]{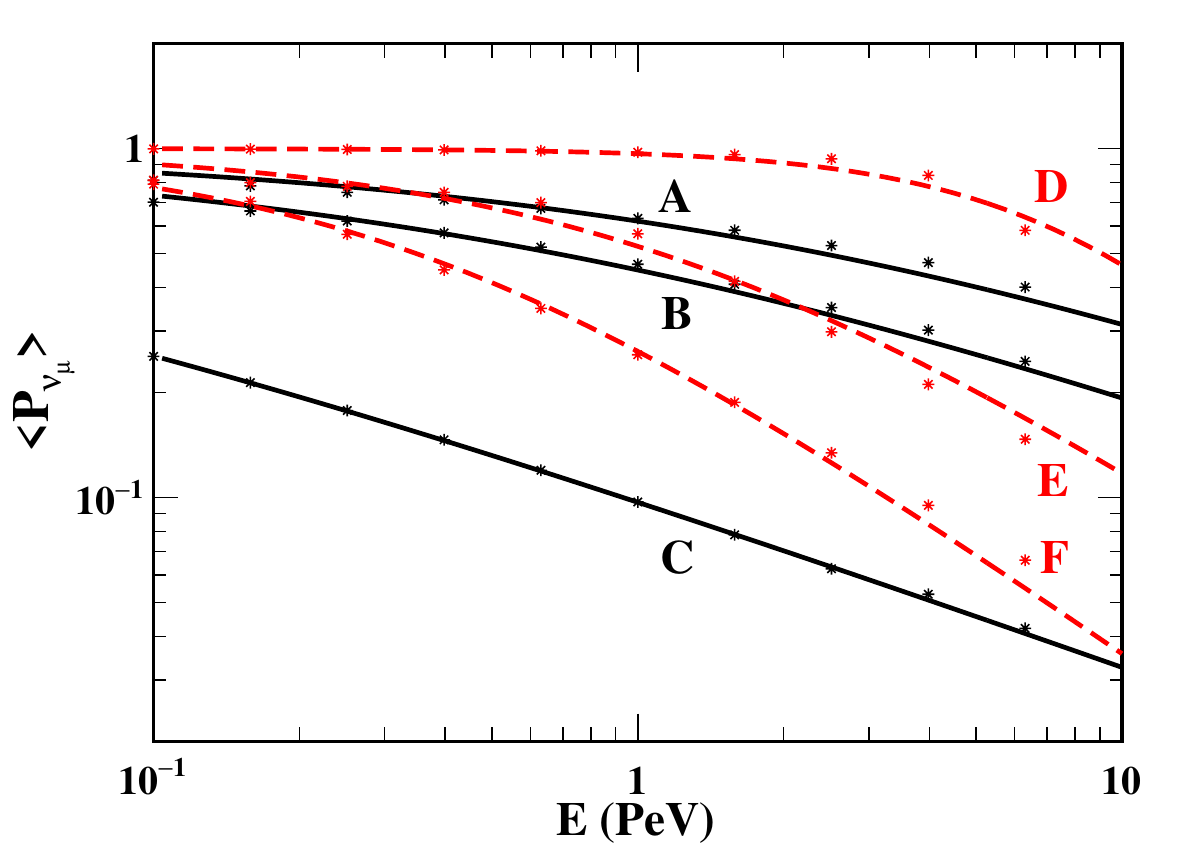}
\caption{The fit of $\langle P_{\nu_\mu}(E)\rangle$ to Eq.~\eqref{eq:pnu} (solid lines), with $\ast$ representing the numerically calculated $\langle P_{\nu_\mu}\rangle$ from Eq.~\eqref{eq:pe} at specific values of $E$ for the NO scenario of LE neutrino flavor evolution. For cases A, B, and C, $g_\Gamma(\theta_0)$ is used, and 
$(T_{\nu,{\rm MeV}},R_{{\rm sh},9},\Gamma_\mathrm{sh})=(3,10,5)$,
$(6,10,5)$, and $(8,3,3)$, respectively. For cases D, E, and F, $g_\theta(\theta_0)$ is used, and 
$(T_{\nu,{\rm MeV}},R_{{\rm sh},9},\theta_\mathrm{sh})=(6,100,10^\circ)$,
$(6,100,20^\circ)$, and $(6,100,30^\circ)$, respectively.}
\label{fig:fit} 
\end{figure}

\begin{figure*}[htbp]
\includegraphics[width=\columnwidth]{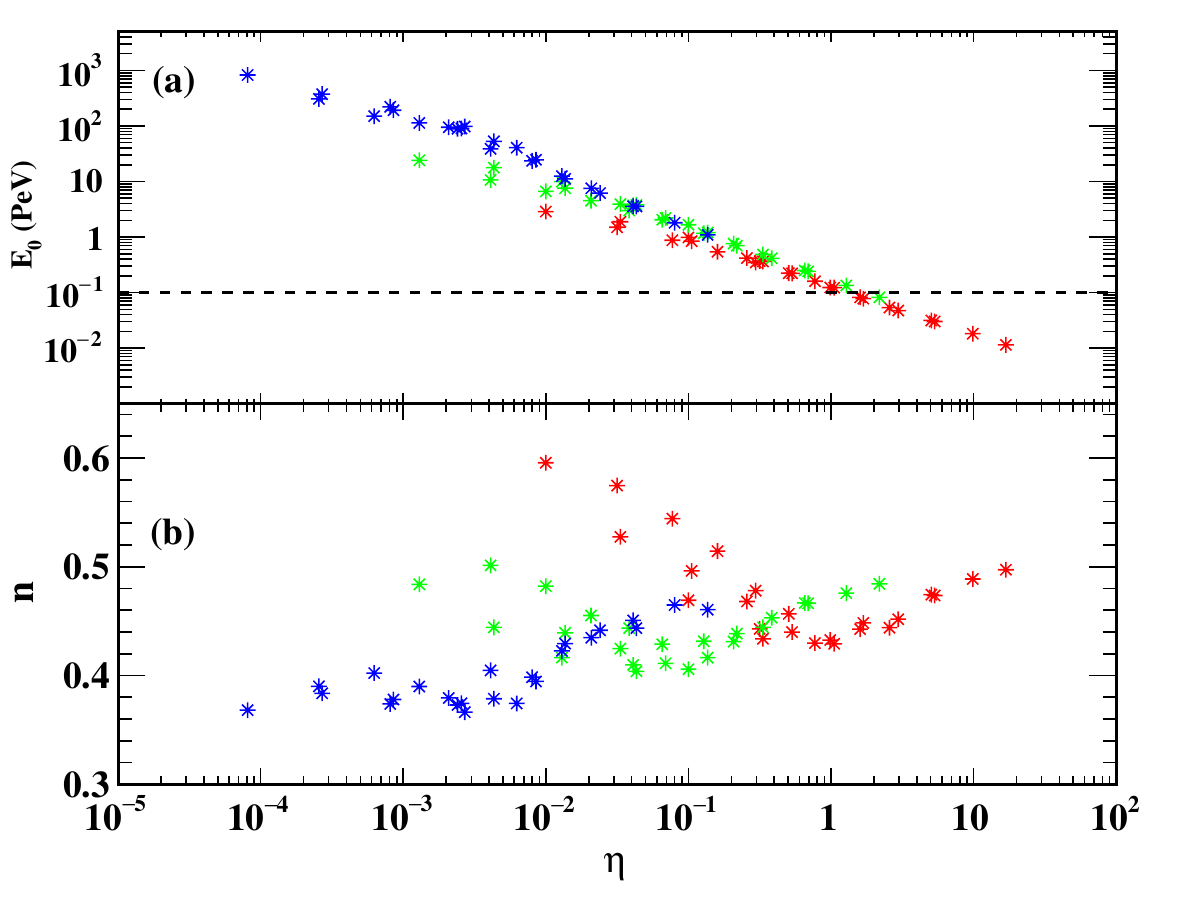} 
\includegraphics[width=\columnwidth]{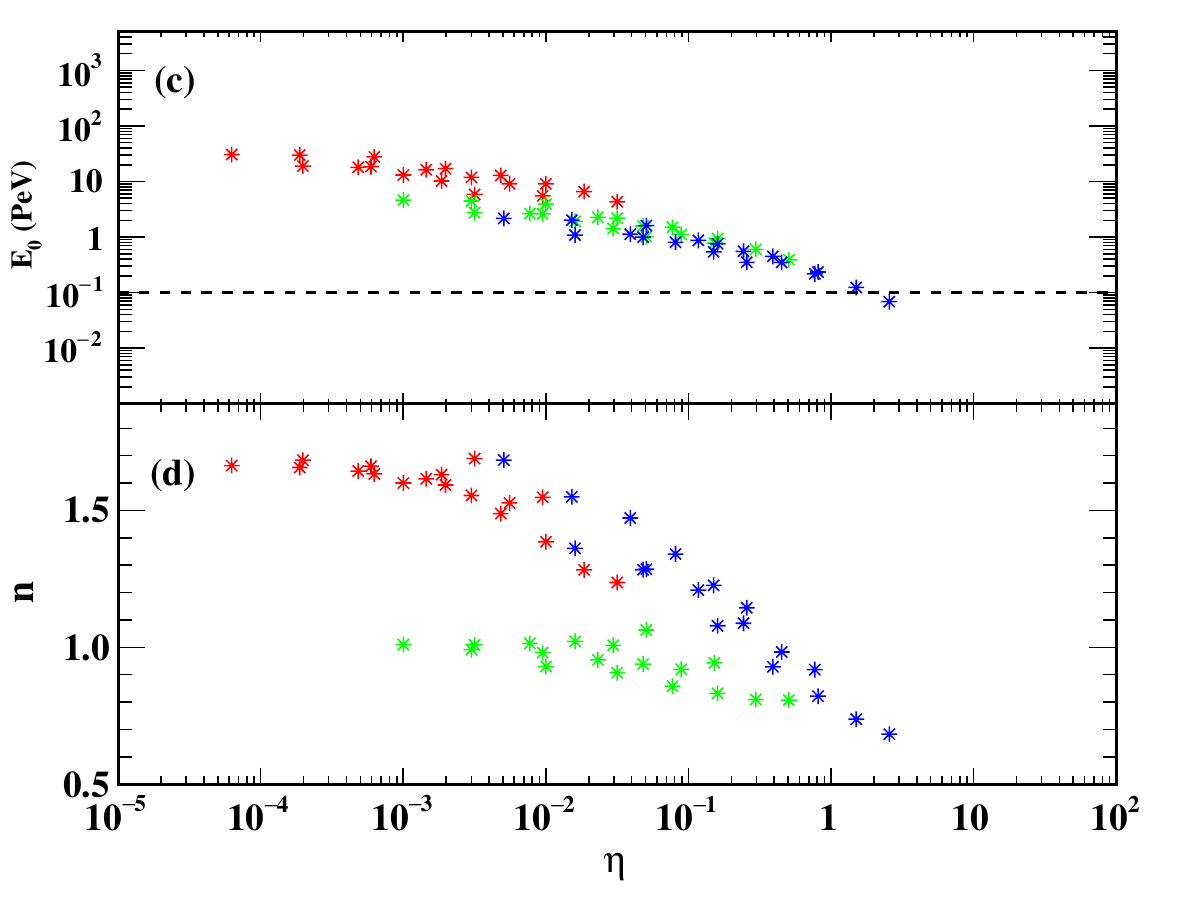} 
\caption{Fitting parameters $E_0$ and $n$ as functions of $\eta$ for 
$\langle P_{\nu_\mu}(E)\rangle$ in the NO scenario of LE neutrino flavor evolution. Left panel: the three
trends for $E_0$ and $n$
are for $\Gamma_\mathrm{sh}=10$ (blue), 5 (green), and 3 (red), 
respectively. Right panel: the three
trends are for $\theta_\mathrm{sh}=10^\circ$ (red), $20^\circ$ (green), and $30^\circ$ (blue), 
respectively. The dashed line corresponds to $E_0=0.1$ PeV.} 
\label{fig:Scale}
\end{figure*}

As we aim to estimate the effect of annihilation on the diffuse HE neutrino flux from similar sources, we average $P_{\nu_\alpha}(E,\theta_0)$ over $\theta_0$ considering both distributions given by Eqs.~\eqref{eq:gem} and \eqref{eq:gem_theta} to obtain
\begin{equation}
\langle P_{\nu_\alpha}(E)\rangle=\int_{-1}^{1}\exp[-\tau_{\nu_\alpha}(E,\theta_0)]
g_{\Gamma, \theta}(\theta_0)d\cos\theta_0.
\label{eq:pe}
\end{equation}
For quantitative estimates, we take 
$R_{\rm sh, 9}=3$, 10, 30, 100, and
$\Gamma_{\rm sh}=3$, 5, 10 when $g_\Gamma(\theta_0)$ is applied. For $g_\theta(\theta_0)$, we consider slightly larger $R_{\rm sh}$ with $R_{\rm sh, 9}=30$, 100, 300 and take $\theta_{\rm sh}=10^\circ$, $20^\circ$, and $30^\circ$. In both cases, we take $R_{\nu,7}=1$, and 
$T_{\nu,{\rm MeV}}=3$, 4, 5, 6, 7, and 8. For all 
these conditions and all five LE neutrino flavor evolution scenarios, we find that 
\begin{equation}
\langle P_{\nu_\alpha}(E)\rangle=[1+(E/E_0)^n]^{-1}
\label{eq:pnu}
\end{equation}
is a good fit over $E=100$~TeV to 10~PeV, as demonstrated by Fig.~\ref{fig:fit} for $\langle P_{\nu_\mu}(E)\rangle$ with selected parameter sets. 
The same form of fit with 
slightly different $E_0$ and $n$ also applies to 
$\langle P_{\bar\nu_\alpha}(E)\rangle$. 

The parameter $E_0$ is a characteristic energy for which annihilation with 
thermal neutrinos is significant. For the case where $\theta_0$ follows the distribution $g_\Gamma(\theta_0)$ with $\theta_0\sim\Gamma_\mathrm{sh}^{-1}$, we expect from 
Eq.~(\ref{eq:tau2}) that $E_0$ should inversely scale with
\begin{equation}
\eta = R_{\nu,7}^2T_{\nu,{\rm MeV}}^4R_{{\rm sh},9}^{-1}\Gamma_\mathrm{sh}^{-4}.
\label{eq:eta}
\end{equation}
Similarly, if $\theta_0 \sim \theta_{\rm sh}$ under the distribution $g_\theta(\theta_0)$, the corresponding $\eta$ is given by \begin{align}
\eta = 0.25 R_{\nu,7}^2T_{\nu,{\rm MeV}}^4R_{{\rm sh},9}^{-1}\theta_\mathrm{sh}^{4}.
\label{eq:eta2}
\end{align}
For illustration, Fig.~\ref{fig:Scale} shows $E_0$ and $n$ as functions of $\eta$ for 
$\langle P_{\nu_\mu}(E)\rangle$ in the NO scenario of LE neutrino flavor evolution. It can be seen that 
$E_{0,{\rm PeV}}\sim 0.1/\eta$ for all combinations 
of $T_\nu$, $R_{\rm sh}$, and $\Gamma_\mathrm{sh}$ or $\theta_{\rm sh}$ considered, in agreement with 
Eq.~(\ref{eq:tau2}). [The effect of $R_\nu$ is as in Eq.~(\ref{eq:tau2}) but
not shown.] The behavior of $n$ is more complex. It clearly scales with 
$\eta$ for fixed $\Gamma_\mathrm{sh}$ or $\theta_{\rm sh}$, but the trend varies with $\Gamma_\mathrm{sh}$ or $\theta_{\rm sh}$. 
However, when annihilation of $1$~PeV neutrinos becomes significant for 
$\eta\gtrsim 0.1$ (see the dashed horizontal line of Fig.~\ref{fig:Scale}), $n$ lies in a narrow range of $\approx 0.4$--0.5 for $g_\Gamma$ and ranges from 0.7--1.3 for $g_\theta$
(see Fig.~\ref{fig:Scale}b and Fig.~\ref{fig:Scale}d).

The explicit dependence of $n$ on $\Gamma_\mathrm{sh}$ can be traced to the contribution
from $\nu_\alpha\bar\nu_\alpha$ annihilation. In contrast to the approximate linear
scaling with $s$ for the cross section of $\nu_\alpha\bar\nu_\beta$ 
($\beta\neq\alpha$) annihilation, the cross section of $\nu_\alpha\bar\nu_\alpha$ 
annihilation has a resonant form
\begin{equation}
\sigma_{\nu_\alpha\bar\nu_\alpha}\sim\frac{G_F^2M_Z^4s}
{(s-M_Z^2)^2+\Gamma_Z^2M_Z^2},
\label{eq:xs}
\end{equation}
where $M_Z$ is the mass of the $Z$ boson and $\Gamma_Z$ is its decay width. 
Taking $E' \sim 3T_\nu$ and $s\sim E'E\theta^2\sim
3T_\nu E \theta_\mathrm{sh}^2
\sim 
6T_\nu E/\Gamma_\mathrm{sh}^2$, one would naively
estimate that the $Z$ resonance occurs for
\begin{equation}
E_{\rm PeV} \sim 3T_{\nu,{\rm MeV}}^{-1}\theta_\mathrm{sh}^{-2} \sim T_{\nu,{\rm MeV}}^{-1}\Gamma_\mathrm{sh}^2.
\label{eq:zres}
\end{equation}
The above estimate indicates that the resonance has little effect on HE neutrinos of PeV and below for
the case of $\Gamma_\mathrm{sh}=10$, but may affect $\langle P_{\nu_\alpha}(E)\rangle$ for $\Gamma_\mathrm{sh}=3$ and 5.
In fact, for the latter cases, the resonance starts to play a role for HE neutrinos with energies lower than that given by Eq.~\eqref{eq:zres}
because the intersection angle $\theta_0$ follows a broader distribution [Eq.~\eqref{eq:gem}] for smaller $\Gamma_\mathrm{sh}$.
The resonance significantly affects $\langle P_{\nu_\alpha}(E)\rangle$ at 
$E\gtrsim 0.1$~PeV and $\gtrsim 1$ PeV for $\Gamma_\mathrm{sh}=3$ and 5, respectively. 
Consequently, $n$ increases with decreasing $\eta$ at $\eta \lesssim 1$ (corresponding to $E_0 \gtrsim 0.1$ PeV) for $\Gamma_\mathrm{sh}=3$ 
and stays approximately constant at $\eta \lesssim 0.1$ (corresponding to $E_0 \gtrsim 1$ PeV) for $\Gamma_\mathrm{sh}=5$.

In the scenario where $\theta_0$ follows the distribution $g_\theta(\theta_0)$, $n$ tends to decrease with $\eta$. With a smaller $\eta$, and correspondingly, a higher $E_0$, the softening of the HE neutrino spectrum will be more dominated by the $Z$-resonance, as the ``optical" depth at energies far from the resonance region is small [Eq.~\eqref{eq:tau2}]. The dominance of the $Z$-resonance leads to a larger $n$ for a smaller $\eta$. When $\eta$ is large enough (i.e., $\eta \gtrsim$ 0.1), HE neutrinos with energy around and above the resonance energy undergo efficient annihilation. Consequently, only HE neutrinos emitted at $\theta_0 \lesssim \theta_E \sim T^{-1/2}_{\nu,{\rm MeV}} E_{\rm PeV}^{-1/2}$ can effectively survive pair annihilation, leading to $\langle P_{\nu_\alpha}(E)\rangle \propto 1-\cos\theta_E \propto \theta_E^2 \propto E^{-1}$. This is consistent with $n\sim 1$ for $\eta \gtrsim 0.1$, as depicted in Fig.~\ref{fig:Scale}d.

\begin{figure*}[htbp]
\includegraphics[width=\columnwidth]{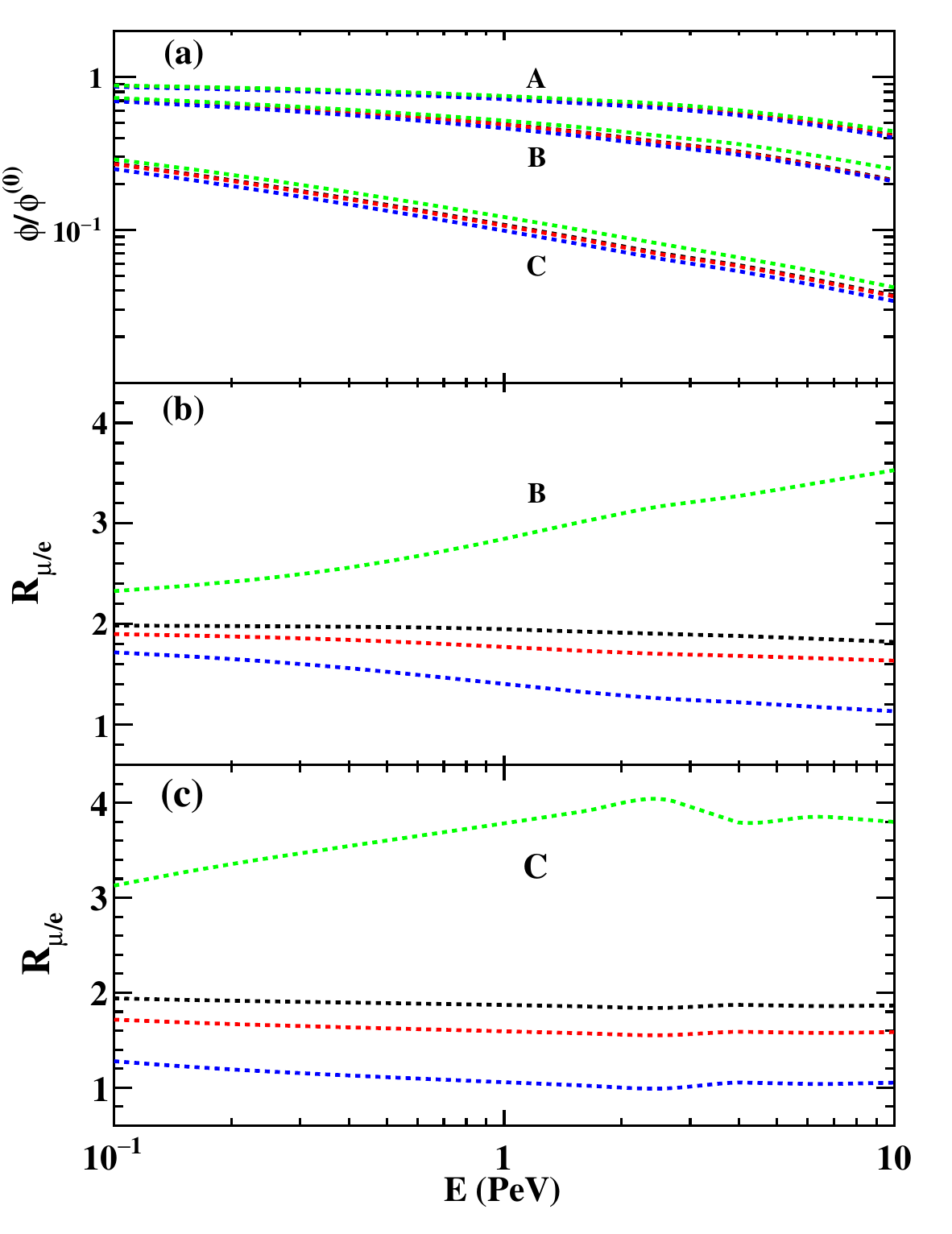}
\includegraphics[width=\columnwidth]{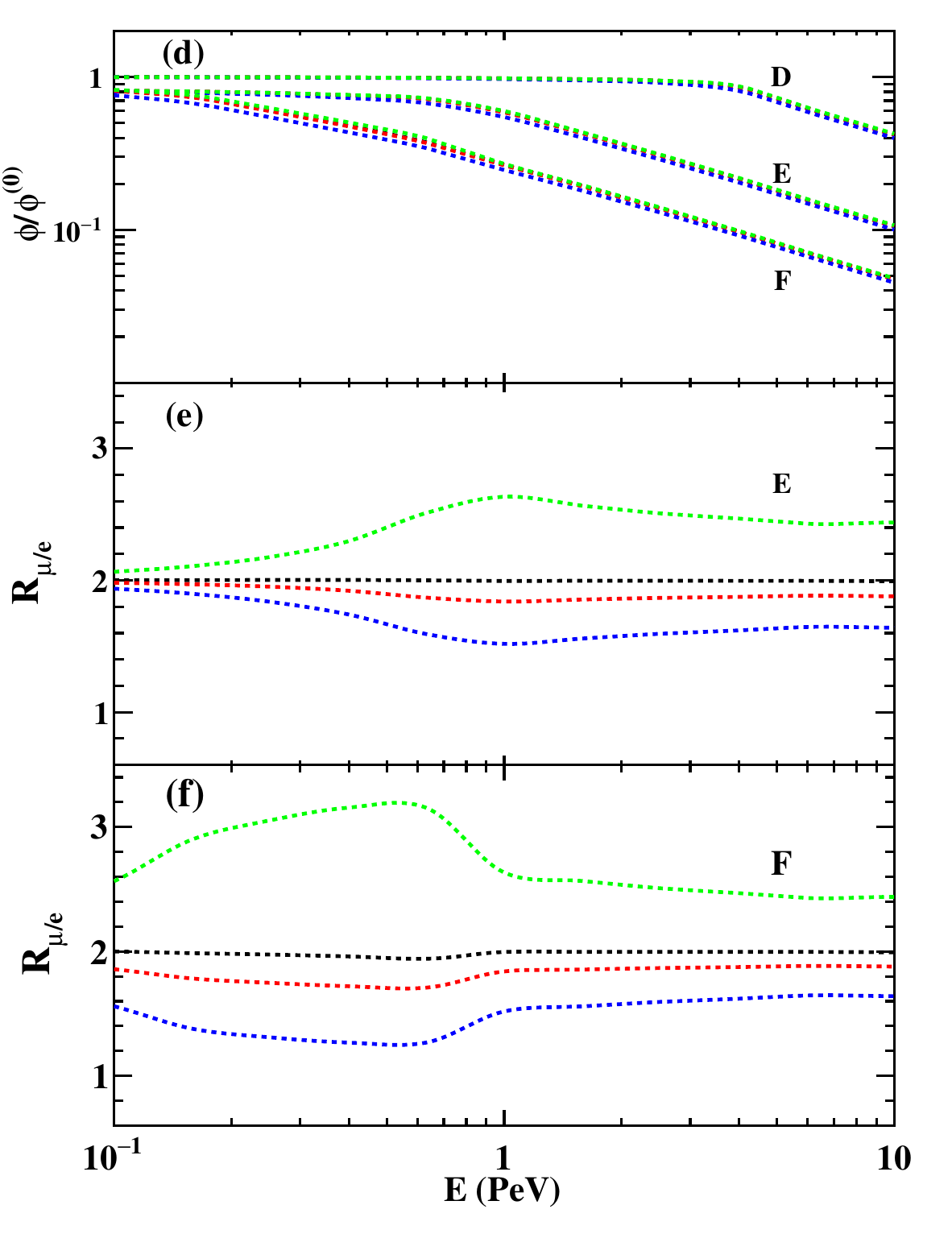}
\caption{Left panel: (a) Effects of $\nu\bar\nu$ annihilation on $\phi/\phi^{(0)}$ for 
$(T_{\nu,{\rm MeV}},R_{{\rm sh},9},\Gamma_\mathrm{sh})=(3,10,5)$,
$(6,10,5)$, and $(8,3,3)$ corresponding to cases A, B, and C, respectively. (b) Effects of $\nu\bar\nu$ annihilation on 
$R_{\mu/e}$ for case B. (c) Same as (b) but for case C.
Right panel: (d) Effects of $\nu\bar\nu$ annihilation on $\phi/\phi^{(0)}$ for 
$(T_{\nu,{\rm MeV}},R_{{\rm sh},9},\theta_\mathrm{sh})=(6,100,10^\circ)$,
$(6,100,20^\circ)$, and $(6,100,30^\circ)$ corresponding to cases D, E, and F, respectively. (e) Effects of $\nu\bar\nu$ annihilation on 
$R_{\mu/e}$ for case E. (f) Same as (e) but for case F. In each case, the curves from top to bottom are for LE neutrino flavor evolution 
scenarios NE, NO, IO, and EE, respectively.}
\label{fig:SpRme}
\end{figure*}

The above results can be used along 
with models of HE neutrino production to estimate signals from a nearby 
source or contributions to the diffuse flux at IceCube.
A proper calculation should include flavor evolution
from the source to IceCube and detailed detector response.
To estimate effects of annihilation with LE neutrinos, we focus on the 
all-flavor spectrum and flavor composition emerging from a source. 
Simply for illustration, we consider 
a case frequently discussed in the literature, where HE $\nu_\mu$, 
$\bar\nu_\mu$, $\nu_e$, and $\bar\nu_e$ are produced initially in ratios of 
$2:2:1:1$ with an all-flavor flux spectrum $\phi^{(0)}(E)$. In this case, the 
emerging all-flavor flux spectrum $\phi(E)$ can be estimated by
\begin{equation}
\frac{\phi}{\phi^{(0)}}\simeq
\frac{\langle P_{\nu_\mu}(E)\rangle+\langle P_{\bar\nu_\mu}(E)\rangle}{3}+
\frac{\langle P_{\nu_e}(E)\rangle+\langle P_{\bar\nu_e}(E)\rangle}{6},
\label{eq:spec}
\end{equation}
and the corresponding flavor ratio is\footnote{It should be pointed out that
Eqs.~\eqref{eq:spec} and \eqref{eq:rmue} neglect 
the secondary HE neutrinos produced from $\nu\bar\nu$ annihilation via the $Z$ resonance.
It is well justified as the branching ratio of the $Z$ boson decay into each neutrino flavor is only $\sim 6$\%.}
\begin{equation}
R_{\mu/e}\simeq\frac{\phi_{\nu_\mu}+\phi_{\bar\nu_\mu}}
{\phi_{\nu_e}+\phi_{\bar\nu_e}}
=\frac{2[\langle P_{\nu_\mu}(E)\rangle+\langle P_{\bar\nu_\mu}(E)\rangle]}
{\langle P_{\nu_e}(E)\rangle+\langle P_{\bar\nu_e}(E)\rangle}.
\label{eq:rmue}
\end{equation}
It is apparent that
Eqs.~\eqref{eq:spec} and \eqref{eq:rmue} can be straightforwardly extended to accommodate other scenarios as well.

The parameters $T_\nu$, $R_{\rm sh}$, $\Gamma_{\rm sh}$, and $\theta_{\rm sh}$ could span relatively wide ranges. To illustrate a significant annihilation effect when $\theta_0$ follows $g_\Gamma(\theta_0)$, we focus on cases with mildly relativistic jets [see Eq.~(\ref{eq:eta})], and take the following three representative parameter sets with $(T_{\nu,{\rm MeV}},R_{{\rm sh},9},\Gamma_\mathrm{sh})=(3,10,5)$, 
$(6,10,5)$, and $(8,3,3)$ for cases A, B, and C, respectively (see also Fig.~\ref{fig:fit}). In the scenario where $\theta_0$ follows the distribution $g_\theta(\theta_0)$, we take $(T_{\nu,{\rm MeV}},R_{{\rm sh},9},\theta_\mathrm{sh})=(6,100,10^\circ)$, 
$(6,100,20^\circ)$, and $(6,100,30^\circ)$ for cases D, E, and F, respectively. The corresponding values of $\eta$ are 0.013, 0.21, 
16.9, 0.003, 0.048, and 0.243 for cases A, B, C, D, E, and F, respectively [see Eqs.~\eqref{eq:eta} and \eqref{eq:eta2}]. We show $\phi/\phi^{(0)}$ as functions of $E$ for different flavor evolution 
scenarios of LE neutrinos in Fig.~\ref{fig:SpRme}a and Fig.~\ref{fig:SpRme}d. For each parameter set, 
$\phi/\phi^{(0)}$ is not sensitive to flavor evolution of LE neutrinos and follows the form of 
$\langle P_{\nu_\alpha}(E)\rangle$ in Eq.~(\ref{eq:pnu}). 
As $E_0$ is $\sim 1$~PeV for $\eta=0.1$ and decreases for larger $\eta$
(Fig.~\ref{fig:Scale}a and Fig.~\ref{fig:Scale}c), annihilation of $\sim 100$~TeV to 1~PeV neutrinos 
is significant for cases B, C, and F. For cases A, D, and E, the effects become important for HE neutrinos above PeV.
In the two most optimistic cases C and F, $\langle P_{\nu_\alpha}(E)\rangle\sim
\langle P_{\bar\nu_\alpha}(E)\rangle\propto E^{-n}$ with $n\approx 0.4$--0.5 and $\approx 1$ 
at $E\gtrsim 0.1$~PeV [Eq.~(\ref{eq:pnu}) and Fig.~\ref{fig:Scale}].

Effects of flavor evolution of LE neutrinos are more evident for $R_{\mu/e}$ as shown in 
Figs.~\ref{fig:SpRme}b, \ref{fig:SpRme}c, \ref{fig:SpRme}e, and \ref{fig:SpRme}f for cases B, C, D, and F, respectively. 
In particular, $R_{\mu/e}$ for scenarios NE and EE differs substantially
from 2 for the case without annihilation of HE neutrinos. Because annihilation 
is more efficient for $\nu$ and $\bar\nu$ of the same flavor, more HE $\nu_e$ 
and $\bar\nu_e$ are annihilated than $\nu_\mu$ and $\bar\nu_\mu$ in 
scenario NE, where thermal $\bar\nu_e$ and $\nu_e$ remain unchanged. This preferential destruction of HE $\nu_e$ and $\bar\nu_e$ will be typically enhanced when 
annihilation probability increases. Across all considered cases, the average survival probabilities initially decrease with $E$, resulting in an increasing $R_{\mu/e}$ with $E$ in scenario NE.

For cases C, E, and F with relatively large $\eta$, the majority of all HE neutrinos emitted at different values of $\theta_0$ would be annihilated, leaving the survival probabilities mainly determined by the fractions of HE neutrinos emitted at small $\theta_0$. In such cases, $\langle P_{\nu_\mu}(E)\rangle$ and $\langle P_{\bar\nu_\mu}(E)\rangle$ tend to get close to $\langle P_{\nu_e}(E)\rangle$ and $\langle P_{\bar\nu_e}(E)\rangle$, indicating a decreasing of $R_{\mu/e}$ towards 2 at high $E$ in scenario NE, as observed in Figs.~\ref{fig:SpRme}c, \ref{fig:SpRme}e, and \ref{fig:SpRme}f. Note that this behaviour is more evident for cases E (Fig.~\ref{fig:SpRme}e) and F (Fig.~\ref{fig:SpRme}f) where the distribution $g_\theta(\theta_0)$ has been applied. In addition, the $Z$-resonance effect starts at
a higher $E$ for the distribution $g_\Gamma(\theta_0)$, which peaks at $\theta_0 =0$. Therefore, $R_{\mu/e}$ for case C in scenario NE starts to decrease at a higher $E$ (Fig.~\ref{fig:SpRme}c). The maximal values of $R_{\mu/e}$ shown in Figs.~\ref{fig:SpRme}b, \ref{fig:SpRme}c, \ref{fig:SpRme}e, and \ref{fig:SpRme}f are 3.5, 4, 2.7, and 3.2 for cases B, C, E and F, respectively.

In scenario EE, all thermal $\bar\nu_e$ ($\nu_e$) are converted into $\bar\nu_\mu$ ($\nu_\mu$). Therefore, the preferential destruction
of HE $\nu_\mu$ ($\bar\nu_\mu$) results in a reduced $R_{\mu/e}$.
In scenarios NO and IO, thermal neutrinos evolve into combinations
of three flavors, for which annihilation of HE $\nu_\mu$ and $\bar\nu_\mu$
is comparable to that of $\nu_e$ and $\bar\nu_e$.
This reduces $R_{\mu/e}$ 
from 2 by significantly smaller amounts than scenario EE 
(Figs.~\ref{fig:SpRme}b, \ref{fig:SpRme}c, \ref{fig:SpRme}e, and \ref{fig:SpRme}f). For scenario FE (not shown), the flavor composition of HE neutrinos is not affected and $R_{\mu/e}$ is always equal to 2.

\section{Discussion and summary}
\label{sec:summary}

We have conducted a parametric and largely model-independent study to investigate the impacts of annihilation with LE neutrinos on HE neutrinos. In this section, we briefly discuss the potential significance of the annihilation effects on HE neutrinos generated by choked jets in massive stars or BNSMs within various models.

The jets, especially those (partly) powered by the annihilation of accretion disk neutrinos, will propagate within the LE neutrino bath. The effects of these LE neutrinos on the jet dynamics and particle acceleration in jet-induced shocks need to be estimated.
Taking a typical cross section of $10^{-42}$~cm$^2$ for LE neutrinos of 10 MeV, the associated optical depth for nucleons within the jet at radius $r$ is $\sim 10^{-6} L_{\nu,53} r_{10}^{-1}$.
However, the interaction between LE neutrinos and HE protons
may become significant during acceleration, as the cross section of the $p\nu$ process increases with the proton energy. It can potentially compete with the $p\gamma$ process and thus have a notable impact on proton acceleration and HE neutrino production. For a proton of 1 PeV, a neutrino of 10 MeV, and a substantial intersection angle between their momenta, the $p\nu$ cross section is about $10^{-34}~{\rm cm}^2$, which is about 6 orders of magnitude lower than that of the $p\gamma$ process. Assuming that a fraction,  $\epsilon_e$, of the jet energy is converted into thermal radiation, the number density of thermal photons in the comoving frame of shocks can be estimated as $n'_\gamma \sim 10^{26}~ [\epsilon_{e,0.1} L_{\rm iso, 52}/(R_{\rm sh, 10}^2 \Gamma_{\rm sh, 0.5}^2)]^{3/4}$~cm$^{-3}$ \cite{Razzaque2004,Razzaque2005}, where $L_{\rm iso}$ is the jet isotropic luminosity, and $\Gamma_\mathrm{sh}$ and $R_\mathrm{sh}$ are the shock Lorentz factor and radius, respectively. For comparison, the LE neutrino density is $n'_\nu \sim 6\times 10^{26}~L_{\nu,53}\Gamma_{\rm sh, 0.5} R_{\rm sh, 10}^{-2}$~cm$^{-3}$.
Under the conditions explored in this study, the $p\nu$ process is always unimportant relative to the $p\gamma$ process and can be safely ignored.

Production of HE neutrinos at internal shocks ($\Gamma_{\rm sh} \sim 3$--10 and $R_{\rm sh} \sim 10^{9}$--$10^{11}$ cm) caused by slow jets inside a star has been extensively studied \cite{Meszaros2001,Razzaque:2003uv,Razzaque2004,Ando:2005xi,Razzaque2005,Bartos:2012sg,Fraija2013,Varela2015,Tamborra2015,Senno2015,Fraija2015,Denton:2017jwk,Denton:2018tdj,He:2018lwb,Gottlieb:2021pzr,Chang:2022hqj,Guarini:2022hry,Bhattacharya:2022btx}.
It was shown that neutrinos from the slow jet model including charm decays could dominate the TeV--PeV neutirno flux observed at IceCube  \cite{Enberg:2008jm,Gandhi:2009qx,Bhattacharya2014} (see also \cite{Bhattacharya:2023mmp,Valtonen-Mattila:2022nej}). Based on the quantitative studies presented in Sec.~\ref{sec:anni} for the case where $\theta_0$ follows $g_\Gamma(\theta_0)$, the resulting HE neutrino spectrum above $\sim 0.1$ PeV could be softened with a spectral index change of 0.4--0.5 due to neutrino pair annihilation. Additionally, the emerging flavor composition can be modified, depending on the oscillation scenarios of LE neutrinos. An accurate measurement of the flavor composition of
HE neutrinos may provide a test of the slow jet model and the annihilation effect.

It was stressed, however, that shocks generated deep inside a star are likely radiation-mediated and particle acceleration at such shocks tends to be inefficient \cite{Levinson:2007rj,Murase2013}. Such radiation constraints are particularly relevant for the slow jet model discussed above. Moreover, the slow jet model has been tightly constrained by IceCube searches \cite{IceCube:2022rlk}. To form a collisionless shock that facilitates efficient particle acceleration, Ref.~\cite{Murase2013} considered low-power and relativistic choked jets in blue supergiants with large shock radius $\gtrsim 10^{11}$~cm, and found they could contribute significantly to the diffuse flux (see also \cite{Carpio:2020app}). Note that the jet luminosity might exhibit a positive correlation with the LE neutrino luminosity. In cases of low-luminosity and long-duration jets, the annihilation effect may be insignificant due to an inadequate flux of LE neutrinos. 

Another intriguing possibility involves choked jets inside an extended wind surrounding
a Wolf-Rayet (WR) star \cite{Nakar:2015tma,Senno2015}. The associated jets could share similar luminosity, Lorentz factor, and duration as the classical GRB jets. HE neutrinos are anticipated either from the internal shocks in the extended wind at $r\gtrsim 10^{11}$ cm, or from the shock breakout from the wind driven by the choked jets at larger radii (see also \cite{Kashiyama:2013}). If the wind density is low enough, collisionless shocks could emerge at the edge of a WR star, supporting both particle acceleration and HE neutrino production \cite{Nakar_2012}. Typically, the emission angle of HE neutrinos from shock breakout can extend to tens of degrees. Similar discussions also apply to the case of jets propagating inside the ejecta from BNSMs. The generation of HE neutrinos from shock breakout driven by jets in BNSMs at $r \sim 10^{11}$ cm has been explored \cite{Gottlieb:2021pzr}. For these models, our parametric study assuming a uniform distribution of $\cos\theta_0$ with $\theta_0\leq\theta_{\rm sh}$ is relevant, indicating a notable steepening in the emerging spectrum and variations in the flavor content of HE neutrinos above a few hundred TeV (Fig.~\ref{fig:SpRme}d, Fig.~\ref{fig:SpRme}e, and Fig.~\ref{fig:SpRme}f).

Although with a lower flux expected, HE neutrinos produced at mildly relativistic collimation shocks inside a star \cite{Murase2013} or the ejecta from BNSMs \cite{Murase:2013hh} could undergo pair annihilation with the LE neutrinos, resulting in further suppression above 0.1 PeV. In cases of highly magnetized jets, particles can also be accelerated through magnetic reconnection close to the stellar center (see, e.g., \cite{Guarini:2022hry}). Pair annihilation could also affect the spectrum and flavor composition of the HE neutrinos produced in these cases.        

We have studied the annihilation of HE and LE neutrinos in BNSMs and
CCSNe and how this process can affect the spectra and flavor ratio of HE neutrinos emerging from these sources.
Assuming that the emission angle $\theta_0$ of HE neutrino is constrained by the Lorentz factor of the shocked fluid or simply follows a uniform distribution of $\cos\theta_0$ within some opening angle, we show that the potential effect can be characterized by a single parameter $\eta$
[Eqs.~\eqref{eq:eta} and \eqref{eq:eta2}], which captures the key dependence on 
astrophysical conditions relevant to HE neutrino production. Annihilation probability increases with $\eta$ 
and starts to be significant for neutrinos of $\gtrsim 0.1$~PeV (1 PeV) at 
$\eta\sim 1$ (0.1). For a specific $\eta$, annihilation probability increases 
with energy, which modifies the emerging spectra. For 
$\eta\gtrsim 0.1$--1, the all-flavor spectrum at $E\gtrsim 0.1$--1~PeV is 
modified by a factor $E^{-n}$ with $n\gtrsim 0.4$--0.5. 
Moreover, we have found that although the flavor evolution of the accretion disk neutrinos does not affect the above modification of the spectral index, it  
can change the emerging flavor composition of HE neutrinos.
Among the five flavor evolution scenarios that we have considered (see Table~\ref{tab:flavor_evol}), scenarios NE and EE are particularly interesting as they can
lead to large increases and decreases, respectively, of the emerging $\mu$-to-$e$ flavor ratio $R_{\mu/e}$
from the canonical values and do so in an energy-dependent manner. The spectral change and the variations and energy dependence of $R_{\mu/e}$ may be tested at the next-generation large neutrino observatories such as IceCube-Gen2 \cite{Ackermann:2017pja,IceCube-Gen2:2020qha,Song:2020nfh,Bhattacharya:2023mmp}. 

Note that the HE neutrinos produced inside stars may also experience absorption by stellar material. This would result in a cutoff in the HE neutrino spectrum, while retaining the same flavor composition. This effect can be ignored for BNSMs and for HE neutrinos produced at $R\gtrsim 10^{11}$~cm. In particular, for Wolf-Rayet stars with the hydrogen envelope stripped off, the stellar radius can be as small as $3\times 10^{10}$ cm \cite{Woosley:2005gy}. However, the absorption effect can be significant for CCSNe if $R_{\rm sh}\lesssim 10^{10}$ cm. It is also possible that the stellar matter is pushed away by the earlier jets \cite{Mizuta:2008ch} or the winds from the proto-neutron star prior to the formation of the black-hole accretion disk, and therefore, the matter absorption effect is suppressed. These and other relevant details should be addressed by specific models or simulations. 
Finally, the central engine driving relativistic jets or winds could be a magnetar~\cite{Bucciantini:2007hy,Bucciantini_2009}, which also emits substantial fluxes of HE neutrinos \cite{Murase:2009pg,Fang:2013vla,Carpio:2020wzg} and LE neutrinos. We expect that the potential presence of $\nu\bar\nu$ annihilation would similarly impact the HE neutrino flux for the magnetar case as well.

\begin{acknowledgments}
We thank the anonymous referee for providing helpful comments and suggestions.
This work was supported in part by the National Natural Science Foundation of China (12205258) and the Natural Science Foundation of Shandong Province, China [ZR2022JQ04 (G.G.)], the US Department of Energy [DE-FG02-87ER40328 (Y.Z.Q.)], and the National Science and Technology Council (No.~111-2628-M-001-003-MY4), the Academia Sinica (No.~AS-CDA-109-M11), and the Physics Division of the National Center for Theoretical Sciences, Taiwan (M.R.W.).
\end{acknowledgments}

%\bibliography{refs}


\end{document}